\documentclass[epj,final]{svjour}
\usepackage{graphics}
\input{epsf}
\begin{document}
\title{Strange attractor simulated on a quantum computer}

\author{M. Terraneo, B. Georgeot and D. L. Shepelyansky}

\institute {Laboratoire de Physique Quantique, UMR 5626 du CNRS, 
Universit\'e Paul Sabatier, F-31062 Toulouse Cedex 4, France}
\titlerunning{Strange attractor simulated on a quantum computer}
\authorrunning{M. Terraneo, B. Georgeot and D. L. Shepelyansky}

\date{Received:}
\abstract{
We show that dissipative classical dynamics converging to a strange attractor 
can be simulated on a quantum computer.  Such quantum computations allow
 to investigate efficiently the small scale structure of
strange attractors, yielding new information inaccessible to classical 
computers.  This opens new possibilities for quantum simulations of
various dissipative processes in nature.}

\date{\today}

\maketitle


Starting from the work of Lorenz \cite{lorenz}, it has been
realized that the dynamics of many various dissipative systems
converges to so-called strange attractors \cite{ruelle}.  These objects are
characterized by fractal dimensions and chaotic unstable dynamics
of individual trajectories
(see e.g. \cite{ott,lichtenberg}). They appear in nature in very different
contexts, including applications to turbulence and weather
forecast \cite{lorenz,ruelle}, molecular dynamics \cite{hoover},
synchronization \cite{pikovsky},
chaotic chemical reactions \cite{belousov},
multimode solid state lasers \cite{roy} and complex dynamics
in ecological systems \cite{ecology,plankton} and physiology 
\cite{glass}.
The efficient numerical simulation 
of such dissipative systems can therefore lead to many important
practical applications.  

Recently, it has been understood that quantum mechanics allows to
perform computations in a fundamentally new way
(see for review e.g. \cite{divi,ekert,steane,preskill,nielsen}).
Indeed, quantum parallelism can enormously accelerate the 
computation and provide new information inaccessible to classical 
computers.  Well-known examples are Shor's factorization algorithm \cite{shor},
which is exponentially faster than any known classical method,
and Grover's search algorithm \cite{grover}, where the gain is polynomial.  
Even if important progress has been achieved during the last years, 
still it is essential to find new areas where quantum processors 
might give access to new information unreachable classically. 
Especially
interesting are applications to dissipative systems with
irreversible dynamics leading to a loss of information.

In this Letter we analyze how classical dissipative dynamics
can be simulated on such quantum processors.
To this aim, we study a simple deterministic model
where dynamics converges to a strange attractor, and show that
it can be efficiently simulated on a quantum computer. 
Even if the dynamics on the
attractor is unstable, dissipative
and irreversible, a realistic quantum computer 
\cite{divi,ekert,steane,preskill,zoller,kane}
can simulate it in a reversible way, and, already with 70 qubits,
will provide access to new informations inaccessible for modern
supercomputers.

To study how a quantum computer can simulate
dissipative dynamics leading to a strange attractor, we choose
the deterministic map given by:
\begin{equation}
\label{map}
\bar{y}=y/2 + x \; \mbox{(mod} \;\mbox{2)}\;\;, \;\; 
\bar{x}=x+\bar{y} \;\mbox{(mod} 
\;\mbox{1)}\;.
\end{equation}
where $-0.5 \leq x < 0.5$, $-1 \leq y < 1$ and bars note 
the new values of variables.
The map has one positive $\lambda_+$ and one negative $\lambda_-$
Lyapunov exponents ($ \lambda_{\pm} = \ln[(5 \pm \sqrt{17})/2]$) 
so that the dynamics converges to a strange attractor 
with Hausdorff ($D_H$) and information ($D_I$)  dimensions 
\cite{kaplan,grassberger,procaccia} 
$D_H \approx D_I = 1+  \lambda_+/|\lambda_-| \approx 1.543 $. 

To implement this map for a computer simulation it is necessary to
discretize the phase space.  We choose the natural discretization in
the binary representation of coordinates $(x,y)$, so that the
dynamics takes place on a regular square lattice with $N \times 2N$ points,
with $N=2^{n_q}$ and $n_q$ integer.  The division in (\ref{map})
is realized by shift
and truncation of the last binary digit. With  this procedure, the 
dissipation generates a discretized irreversible map, displaying a discretized
strange attractor (see Fig.1 top) which approaches the continuous one 
for large $N$.
Such a map can be implemented efficiently on a quantum computer.  
For that, an initial image with $N_d$ points is coded in the wave function 
$|\Psi_0>=\sum_{i,j} a_{ij} |x_i> |y_j>|0>|0>$, where 
$a_{i,j}= 0$ or $1/\sqrt{N_d}$.  Here the two registers $|x_i>$ and $|y_j>$
with $n_q$ and $n_q+1$ qubits 
hold the values of the coordinates $x$ and $y$ 
($x_i= -0.5+i/N, i=0,...,N-1$ and $y_j= -1+j/N, j=0,...,2N-1$).
The third register with $n_q-1$ qubits
is used as workspace for modular additions, and the last one collects
the truncated last digits generated by the divisions (``garbage'').  
We start with the simplified algorithm for which the garbage
gets one digit at each map iteration so that $t$ iterations need
$t$ qubits in the fourth register.  The size of this register
can be significantly reduced using a more refined algorithm
we will describe later.

\begin{figure}[h!]
\centerline{\epsfxsize=4.2cm\epsffile{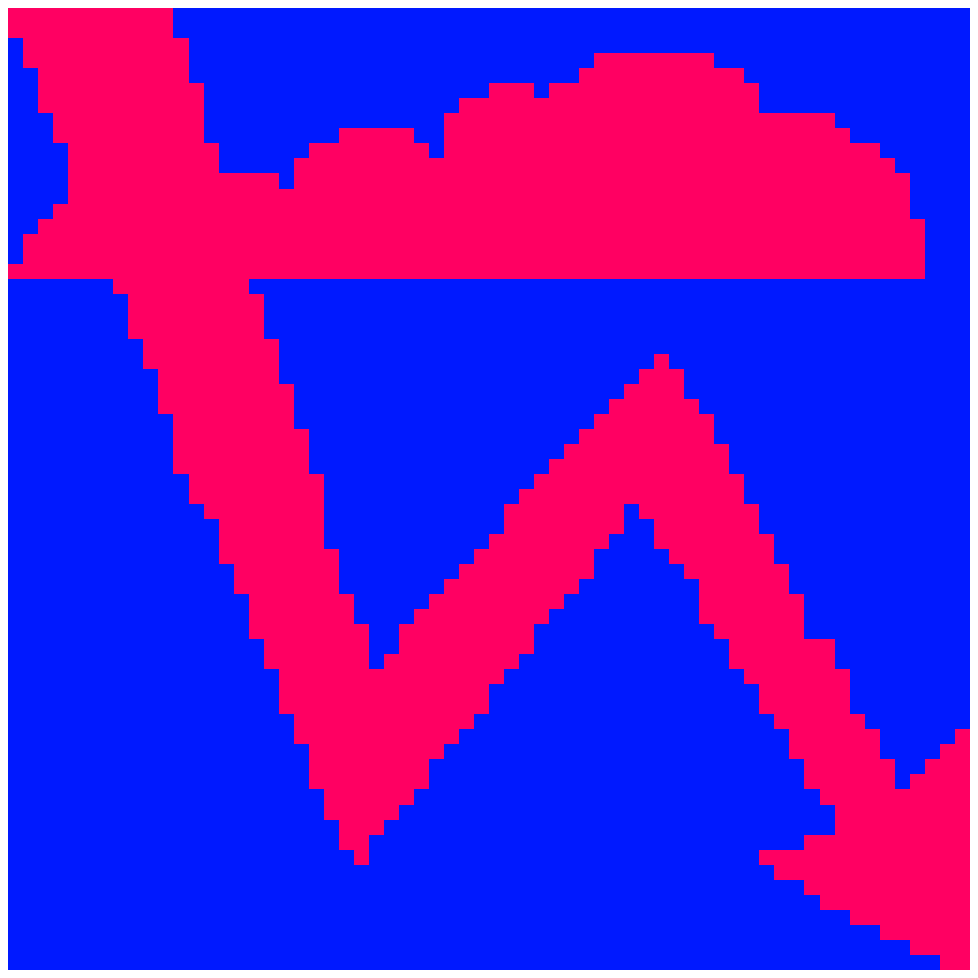}
\hfill\epsfxsize=4.2cm\epsffile{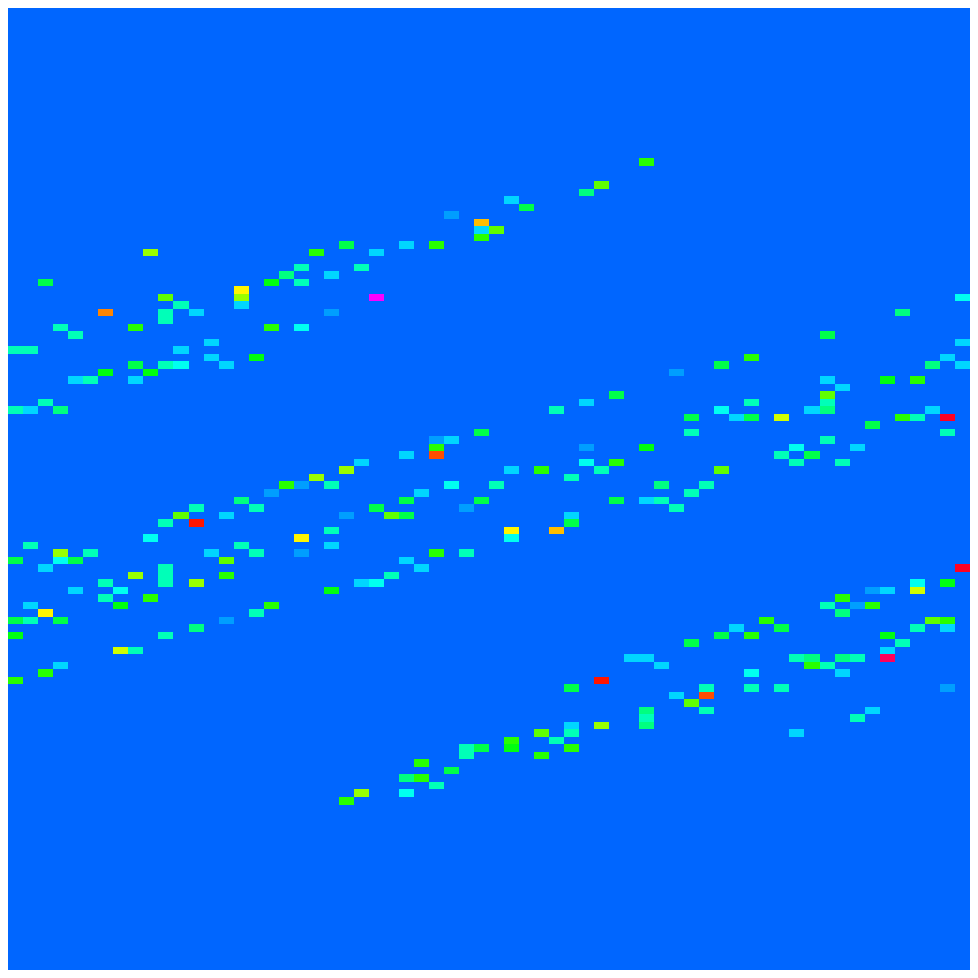}}
\centerline{\epsfxsize=4.2cm\epsffile{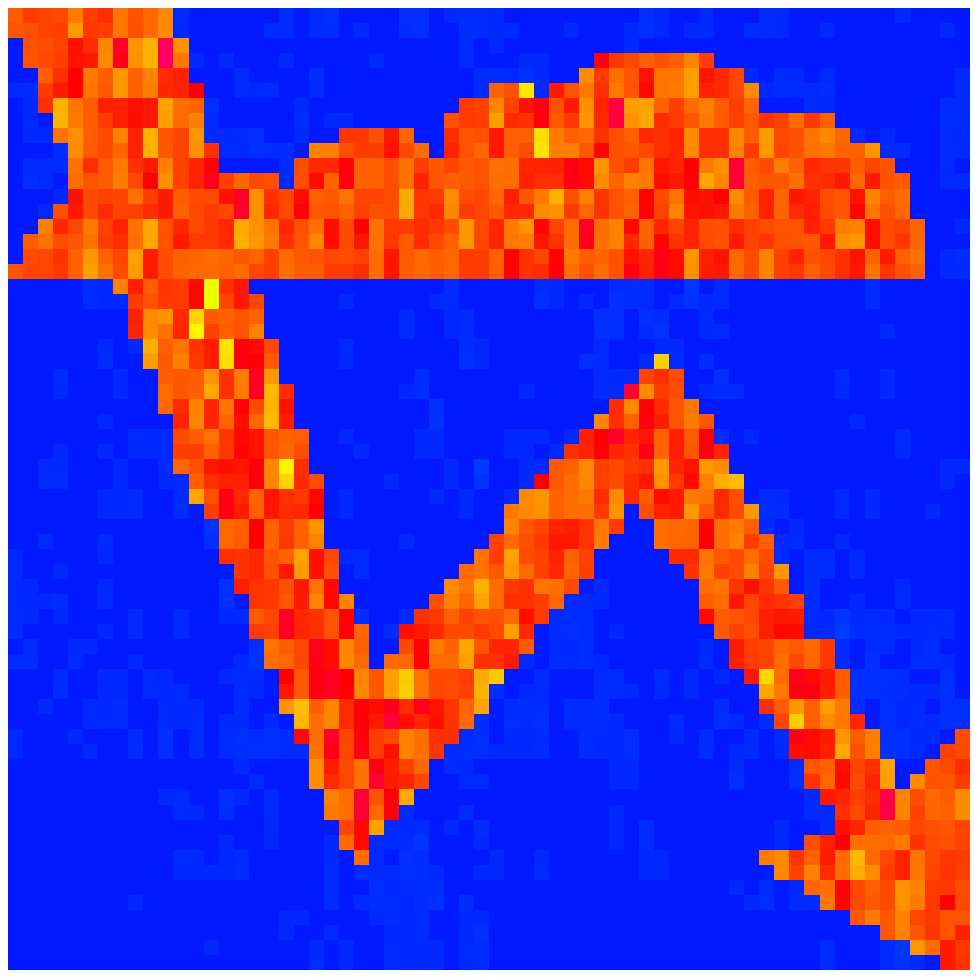}
\hfill\epsfxsize=4.2cm\epsffile{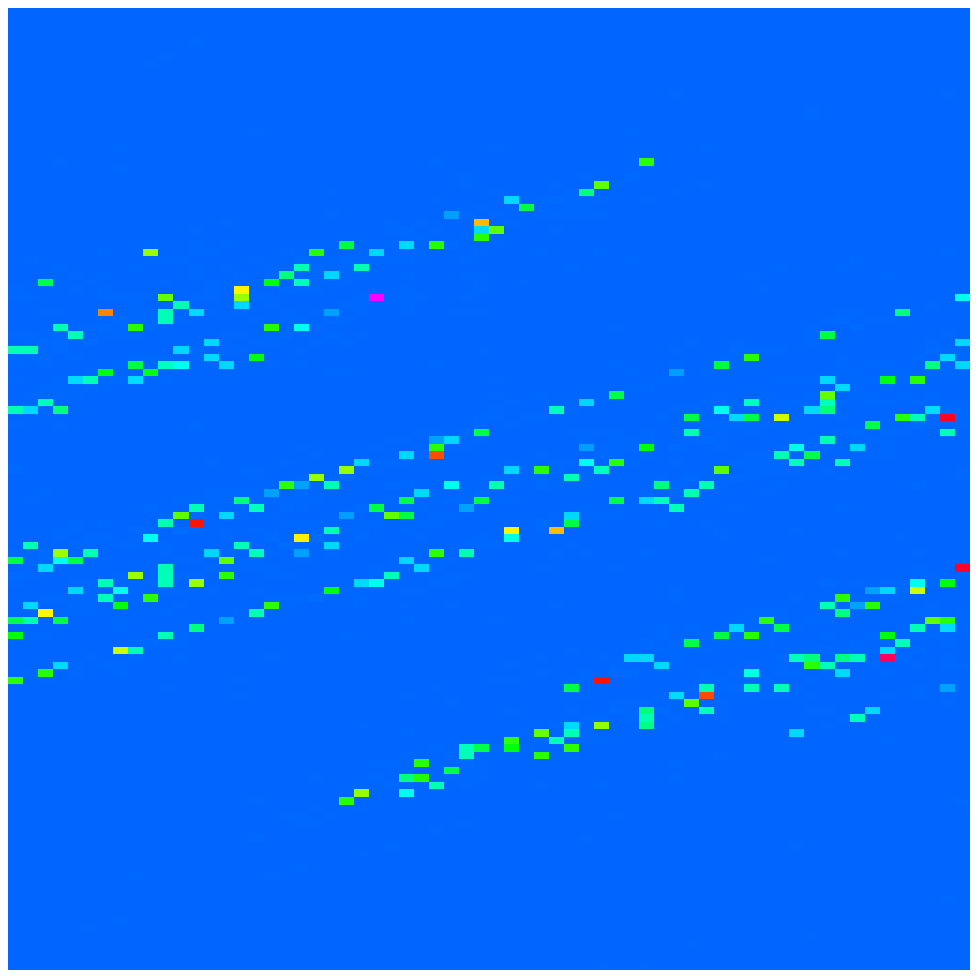}} 
\centerline{\epsfxsize=4.2cm\epsffile{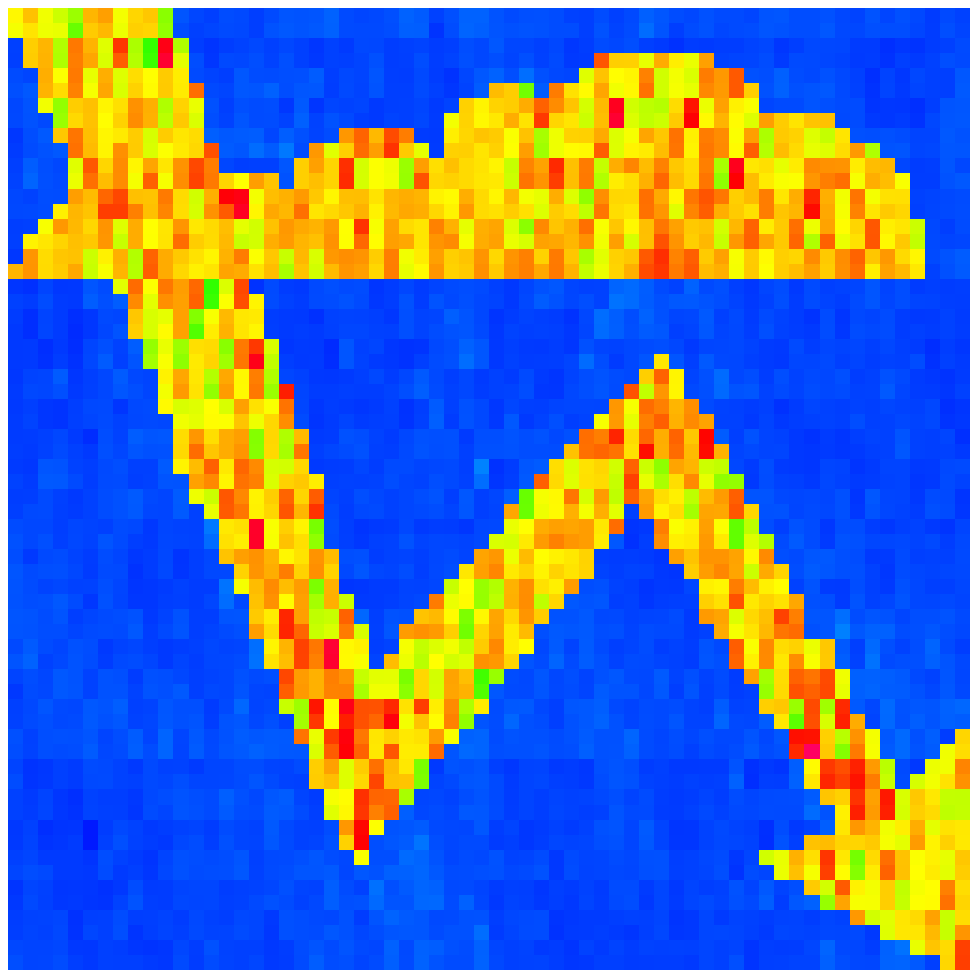}
\hfill\epsfxsize=4.2cm\epsffile{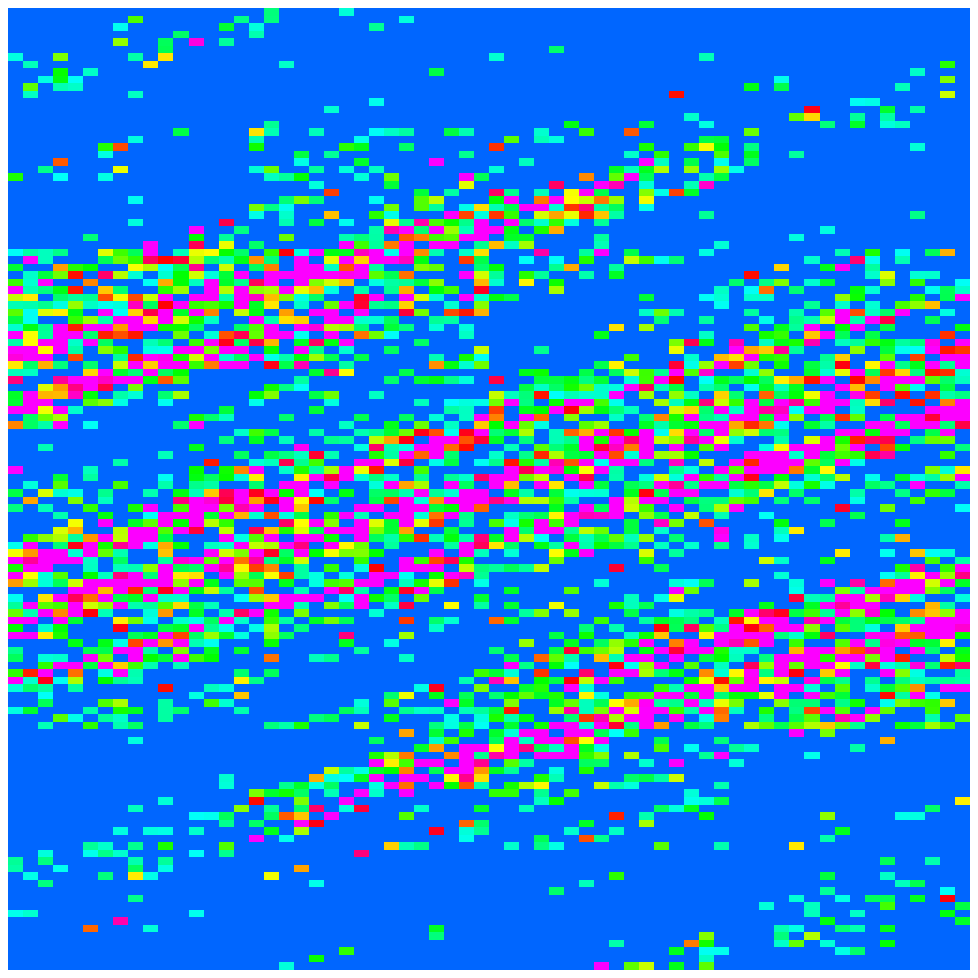}}
\caption{(color) Top: exact classical/quantum computation of the discretized
map (\ref{map});
initial image (left) converged to the
strange attractor after $t=10$ iterations (right).  Middle: 
quantum computation with noise amplitude $\epsilon=0.05$ in each gate
operation; attractor at $t=10$ (right) and initial image recovered
after $10$ backwards iterations with fidelity $f=0.63$ (left). 
Bottom: same as middle with
$\epsilon=0.1$ and $f=0.15$. Left shows the central cell 
($-0.5\leq x,y <0.5$), right shows 
the whole phase space.
Color marks the probability density (integrated over third and fourth 
registers),
 from blue (density less than $10^{-5}$) to red (maximal value). Here
$n_q=6$, with in total $28$ qubits used.
}
\label{fig1}       
\end{figure}
\begin{figure}[h!]
\centerline{\epsfxsize=4.2cm\epsffile{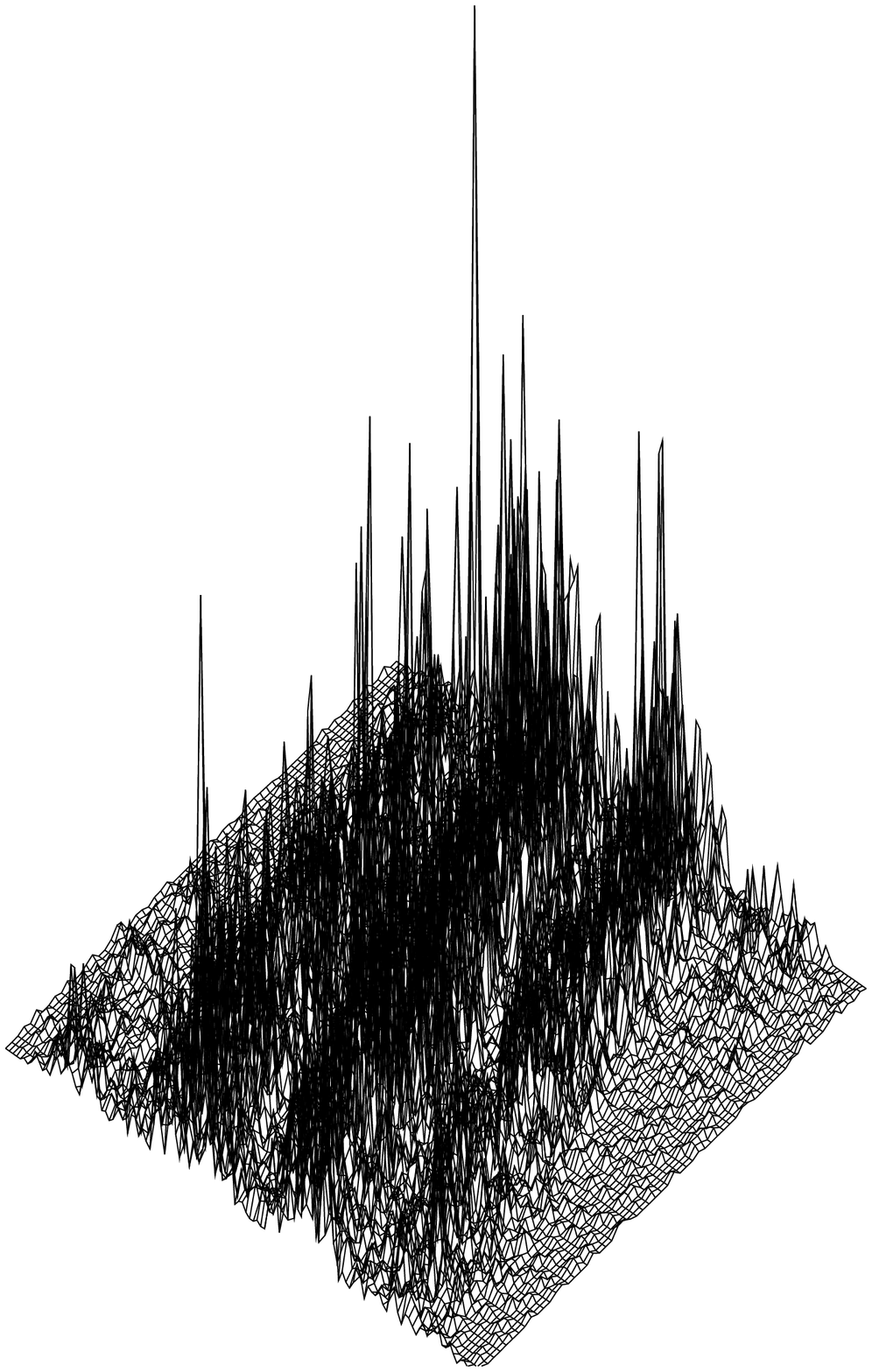}
\hfill\epsfxsize=4.2cm\epsffile{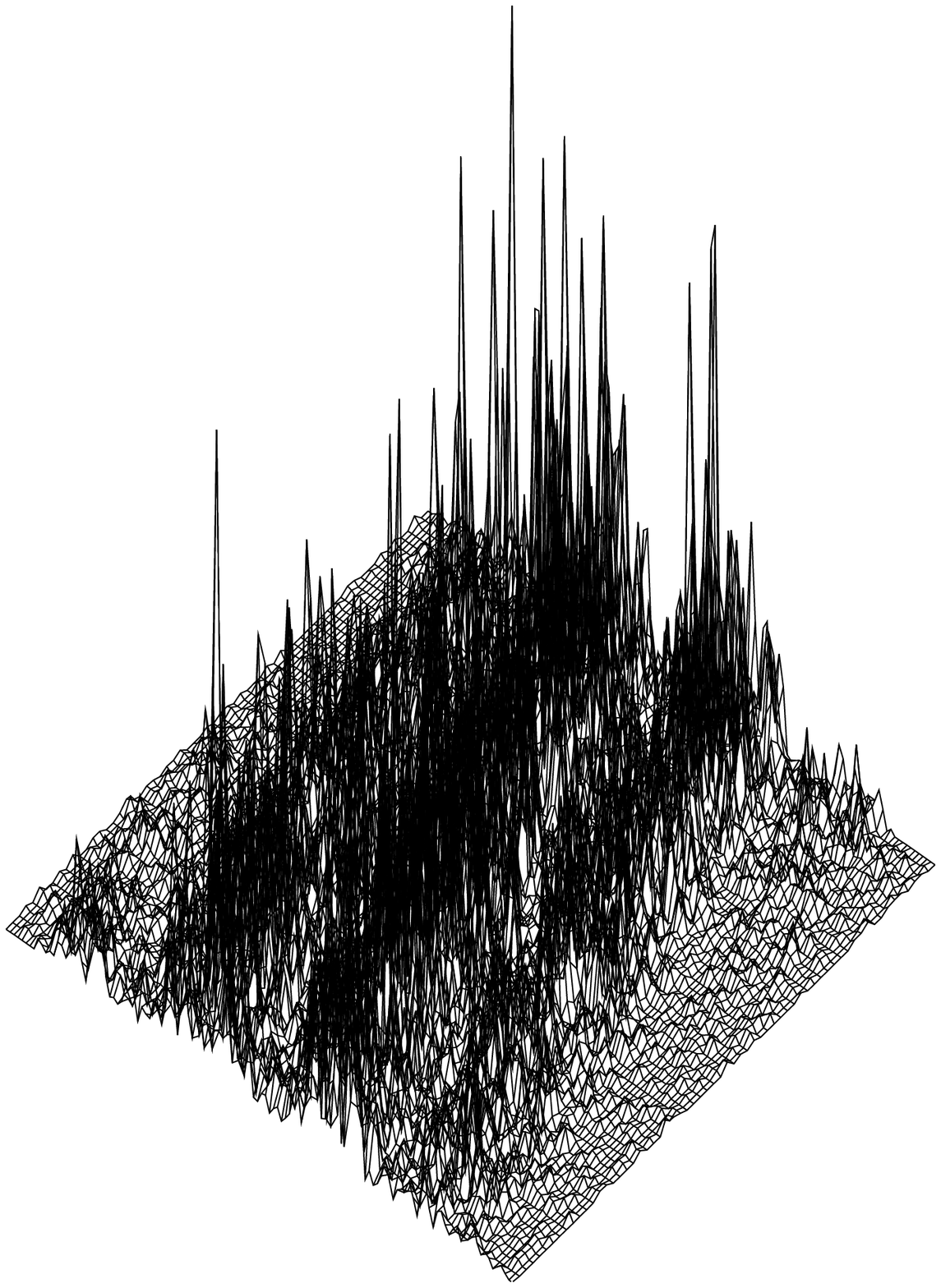}}
\centerline{\epsfxsize=4.2cm\epsffile{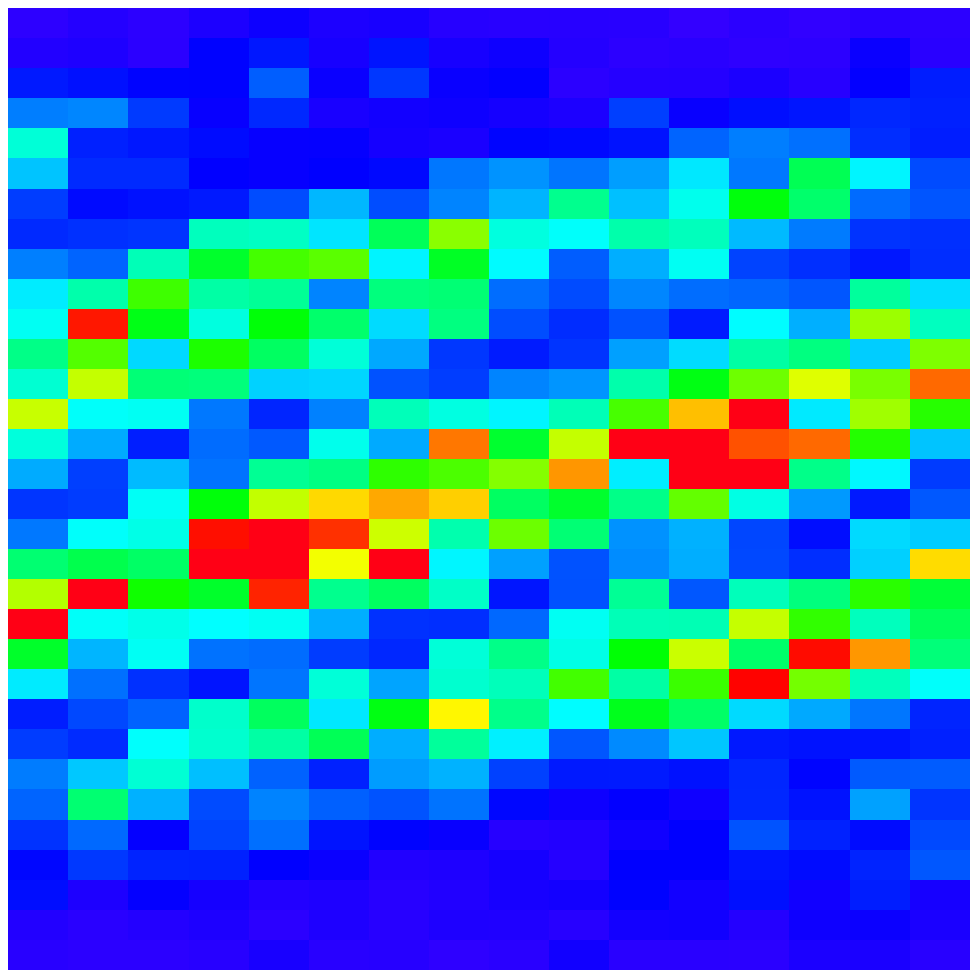}
\hfill\epsfxsize=4.2cm\epsffile{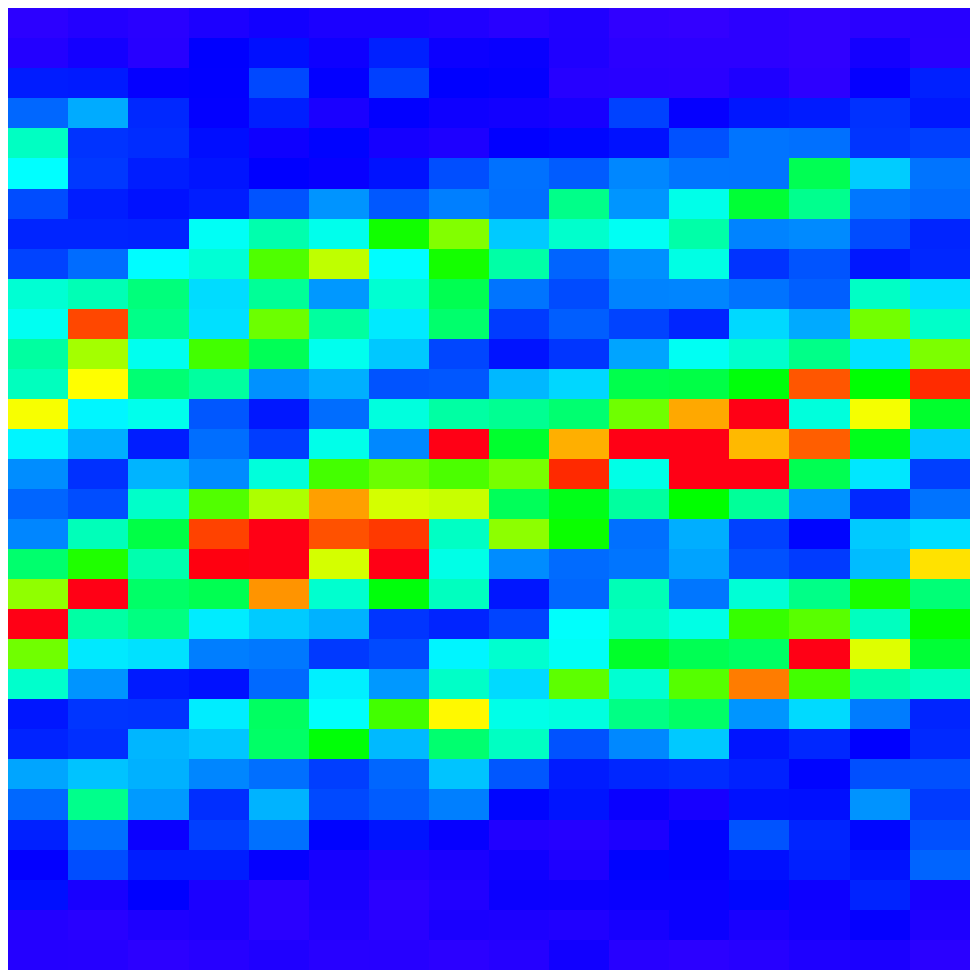}} 
\centerline{\epsfxsize=4.2cm\epsffile{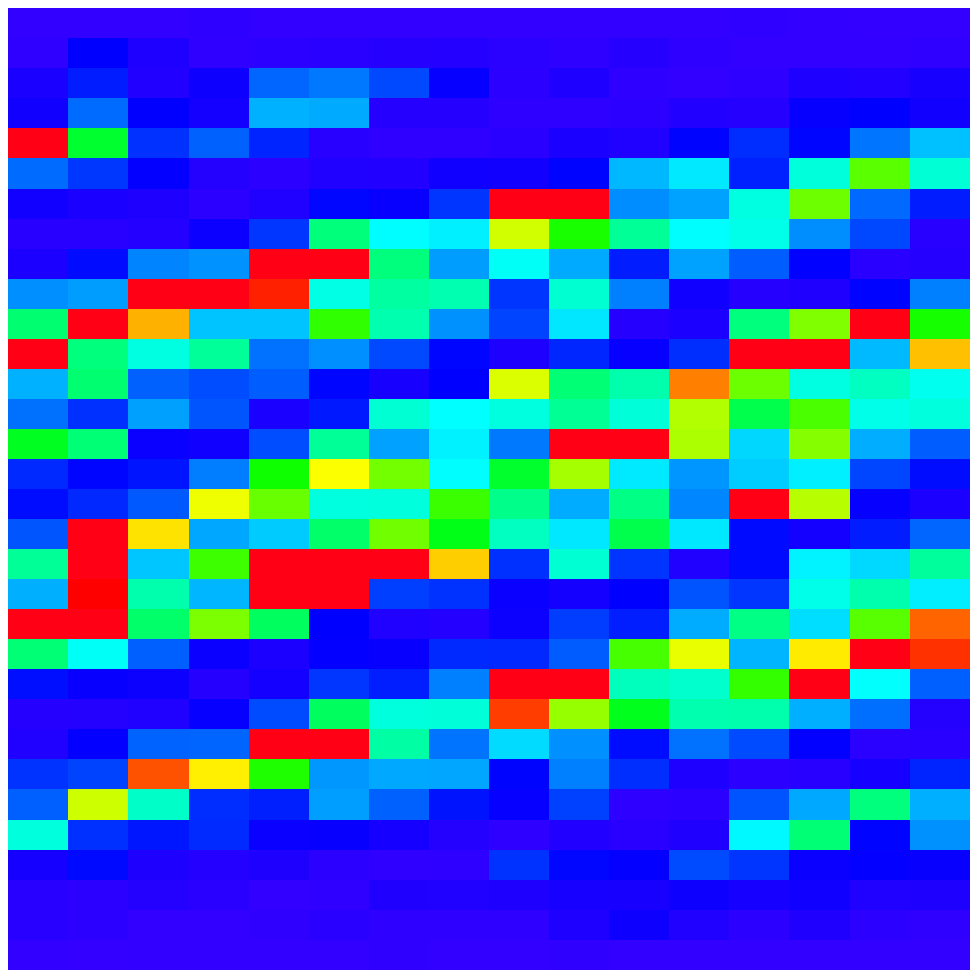}
\hfill\epsfxsize=4.2cm\epsffile{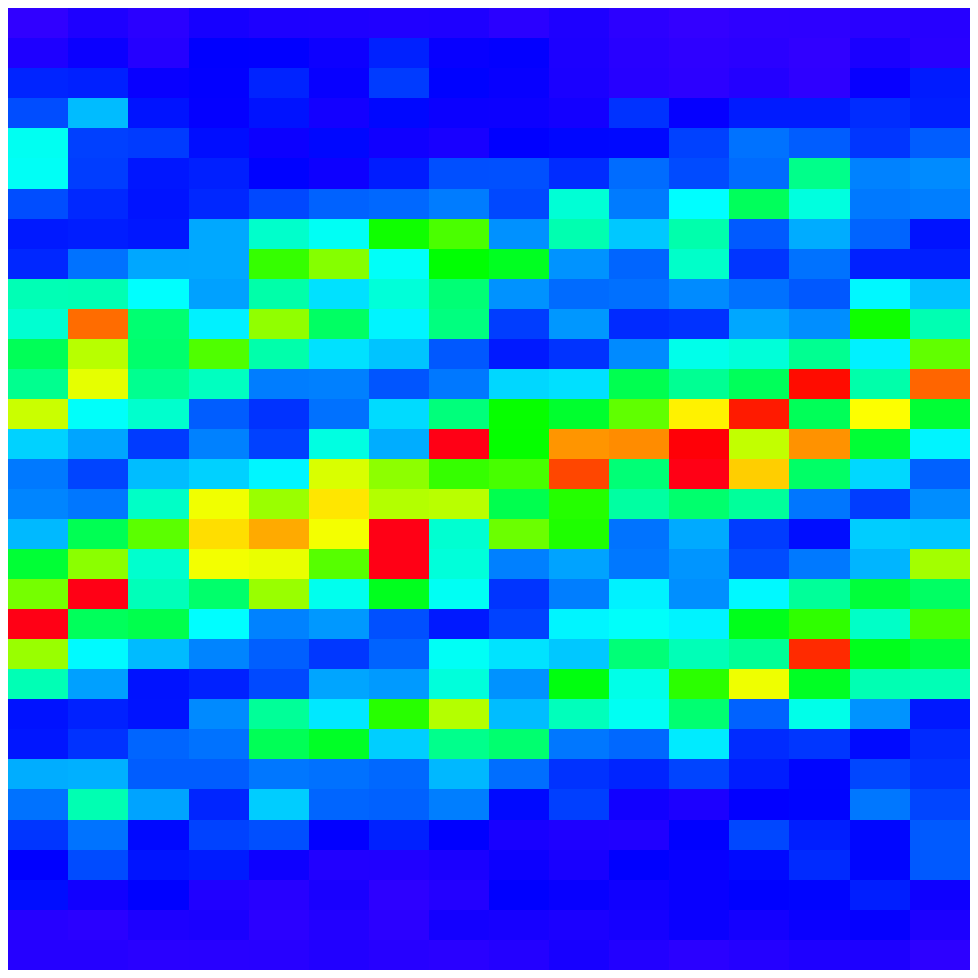}}
\caption{(color) 
Spectral density $|C(t,k_{x,y})|^2$ for the strange attractor
of Fig.1 at $t=10$ in the region 
$-N/2 \leq k_x \leq N/2$, $-N \leq k_y \leq N$ ($N=2^{n_q}$). 
Top: three-dimensional plot of the full 
distribution for $n_q=6$ and  $\epsilon =0$ (left), $\epsilon =0.025$
(right).
Middle: color plot of the coarse-grained distribution  
with $n_q=6$, $n_f=4$ and  $\epsilon =0$ (left), $\epsilon =0.025$
(right).
Bottom: same as middle for $n_q=10$, $\epsilon =0$ (left),
and $n_q=6$, $\epsilon =0.05$ (right).
Colors are as in Fig.1.
}
\label{fig2}       
\end{figure}

The algorithm starts with the initial state $|\Psi_0>$; first it 
places the last qubit of $|y>$ in the garbage register, and uses
$n_q$ swap gates to shift the $|y>$ qubits and obtain
$y/2$. Then a modular addition is implemented in the way described in
\cite{barenco} to add $x$ to $y/2$.  After that, another modular
addition adds the second register to the first.  
In total, this requires $17n_q -10$ quantum operations
 using Toffoli, control-not and swap gates 
in contrast to $O(2^{2n_q})$ operations
for the classical algorithm.
It is interesting to note that the algorithm allows to restore the initial 
state: inverse map iterations are performed ($x=\bar{x}-\bar{y}$,
$y/2=\bar{y}-x$) and $y$ is restored from $y/2$ using a qubit stored in
 the garbage register. This requires a
similar number of operations as the forward iterations.
In principle this can be done on a classical
computer in $O(2^{2n_q})$ operations,
but this requires additional exponentially large  
memory which stores about $2^{2n_q} t$ bits for $N_d \sim N^2$, contrary
to only $t$ qubits used by the quantum computer.

Fig.1  shows the dynamics generated by
the discretized map (\ref{map}) simulated on a quantum computer
with exact and noisy unitary gates with imprecisions of amplitude $\epsilon$.
Due to the dissipative nature of the map (\ref{map}), the initial image
rapidly converges towards the strange attractor (already $t=5$ is enough). 
Even in the presence
of relatively strong noise, the fractal structure of the attractor is 
well-preserved and the initial image can be reliably recovered 
after backwards iterations, despite the exponential instability of
the classical dynamics (\ref{map}).
The precision of computation can be quantitatively characterized through
the fidelity $f$ defined at a given moment of time as 
the projection of the quantum state in presence of gate imperfections
on the exact state without imperfections.  The global properties
of the initial image can be recovered even at relatively low fidelity values.

Even if the quantum algorithm performs one map iteration only in $O(n_q)$
operations, it is important to take into account the measurement procedure
that allows to extract efficiently the information coded in the
wave function.   Indeed the number of points in Fig.1 grows exponentially
with $n_q$ and an exponential number of measurements is required
to obtain the full density distribution.  However, certain characteristics
can be extracted in a polynomial number of measurements, providing new 
information inaccessible for classical computation.  An example of
such a quantity is the spectrum of phase space correlation functions,
defined as 
$C(t,k_{x,y}) = \sum_{x_0,y_0} \exp(2i\pi (x(t,x_0,y_0) + y(t,x_0,y_0))) 
\exp(2i\pi (k_x x_0 +k_y y_0))$, where the sum runs over the points
$(x_0,y_0)$ of the initial distribution, and $(x(t,x_0,y_0) , y(t,x_0,y_0))$
is the position of $(x_0,y_0)$ at time $t$.  Such correlation functions 
have been studied for chaotic systems, where they
determine various kinetic coefficients, for example 
the diffusion rate (see e.g. \cite{lichtenberg} p.328).
Due to chaos, the function $C(t,k_{x,y})$ has significant values
at exponentially high harmonics $k_{x,y} \sim \exp(|\lambda_-|t)$
which rapidly reaches harmonics of order $N$.  
In the theory of classical chaotic dynamics it is well-known that
the information about such
harmonics is very hard to access, since exponentially small
scales should be explored, which can be done only with
exponentially many trajectories \cite{lichtenberg,peres}.  
On the contrary, the quantum computation 
of $C(t,k_{x,y})$ can be done efficiently.  For that, one makes $t$
iterations of (\ref{map}), and creates the state
$\sum a_{x,y} \exp(2i\pi (x + y))  |x> |y>|0>|g(t)>$.  The preparation
of this state is easily done by applying $2n_q +1$ one-qubit rotations
to the first two registers.  Then the garbage $g(t)$ is erased by iterating
the map backwards $t$ times, that at the same time returns the
coefficients $a_{x,y}$ to their original values.  
This creates the state 
$\sum a_{x_0,y_0} \exp(2i\pi(x(t,x_0,y_0) + y(t,x_0,y_0))|x_0> |y_0>|0>|0>$,
keeping phases unchanged.
The whole procedure is
sometimes called ``phase kickback''.  
After that the application
of a two-dimensional quantum Fourier transform \cite{ekert}
yields in $O(n_q^2)$ operations the state 
$\sum C(t,k_{x,y}) |k_x> |k_y>|0>|0>$.  A polynomial number of measurements
yields the principal peaks, or enables
to obtain a coarse-grained image of the
spectral density $|C(t,k_{x,y})|^2$ in the Fourier space. Indeed, 
independently of $n_q$,
one can measure the first $n_f$ and $n_f +1$ 
qubits of the $x$ and $y$ registers respectively,
that gives integrated probability inside $2^{2n_f+1}$ cells.
Fig.2 displays the spectral density for the case of Fig.1.  It
shows that new information about the 
coarse-grained spectral density can be obtained
efficiently.  Indeed, patterns are clearly present in Fig.2 
and they  vary irregularly with $n_q$  
(compare Fig.2 middle left and bottom left).  
This confirms the nontrivial nature of information
provided by the coarse-grained density $|C(t,k_{x,y})|^2$.
Although the spectral density is more sensitive to noise than 
the distribution in Fig.1 still the patterns remain well-defined
even in the presence of relatively strong errors (see Fig.2).

It is important to stress that even
modern supercomputers are unable to find the properties
of the spectral density for $n_q \geq 20$.  Indeed, as is shown
in Fig.3, a classical Monte Carlo algorithm requires
an exponentially large number of trajectories $M$ ($M=O(2^{2n_q})$) to
obtain the coarse-grained spectral density at fixed $n_f$ with fixed accuracy.
In contrast, the quantum computation requires a number of measurements $M$
independent of $n_q$ (each measurement is done after $t$
map iterations and one Fourier transform which needs $O(n_q^2)$ quantum gates).

\begin{figure}
\epsfxsize=3.2in
\epsfysize=2.6in
\epsffile{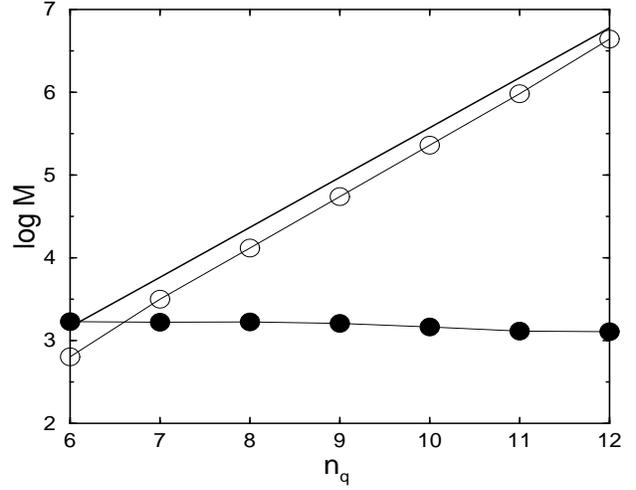}
\caption{Complexity of classical and quantum computations
of coarse-grained distribution of Fig.2 with 
10\% accuracy (\emph{i.e.} fidelity $f(t)=0.9$).
 Here   $n_f=4, t=10$ and different values of $n_q$ are shown. 
For classical computation (open circles),
$M$ gives the number of classical Monte Carlo trajectories
required to reach 10\% accuracy. For quantum case (full circles),
$M$ is the number of measurements
needed to obtain the same accuracy. Full line 
shows the total number of classical points in the initial
image shown in Fig.1 ($M \approx 0.36 \times 2^{2n_q}$).
Logarithm is decimal.
}
\label{fig3}
\end{figure}

To study the effect of noisy gates in a more quantitative way,
we show in Fig.4 the dependence of the fidelity
$f(t)$ of the quantum computation of spectral amplitudes $C(t,k_{x,y})$
on the noise amplitude $\epsilon$ and total number of gates $n_g$ applied
($n_g = t(44 n_q -14 )+ (n_q+2)^2-2$).  The data show the global scaling
law $1-f(t) \approx \epsilon^2 n_g/14$, valid for moderate $\epsilon$. 
The physical origin of this scaling law is related to the fact
that for randomly fluctuating unitary gates the loss of probability
from the exact state is of the order of $\epsilon^2$ for each gate 
operation.  This law determines a time scale 
$t_f \approx 1/(6 \epsilon^2 n_q)$ up to which a reliable quantum 
computation is possible ($f \geq 0.5$).  Beyond $t_f$ the decoherence
destroys the accuracy of the quantum computation and the results
become strongly distorted, see e.g. Fig.2 bottom right. 
We note that a similar time scale appears in quantum
computation of Shor's algorithm on a realistic quantum computer
\cite{zurek}. The computation beyond the scale
$t_f$ is possible but requires the application of quantum 
error-correcting codes, at
the cost of additional qubits and gates
(see e.g. \cite{steane,preskill} and refs. therein).
Without error correction, it is still
possible to improve the fidelity of the final state by 
measuring the third and fourth registers.  Indeed for
exact computation they are at zero after the backwards iterations
while with noisy gates the error probability grows
as $W_g \approx 3.5 \epsilon^2 t$ (Fig.4 inset).  The measurements of
these registers allow to select the correct states and
increase the fidelity by a factor $1/(1-W_g)$, e.g. 
for the case of  Fig.1 bottom this procedure gives 
$f=0.22$.

\begin{figure}
\epsfxsize=3.2in
\epsfysize=2.6in
\epsffile{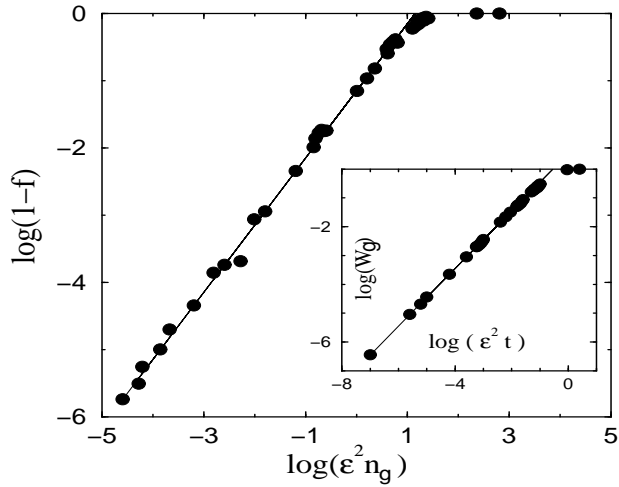}
\caption{Fidelity $f$ for the spectral amplitudes of Fig.2
as a function of $\epsilon^2 n_g$ for $4 \leq n_q \leq 6$,
$10^{-4} \leq \epsilon \leq 0.5$ and $6 \leq t \leq 10$.
Inset: error probability $W_g$ in the third and 
fourth registers as a function of $\epsilon^2 t$ for $n_q=6$.
Straight lines show the theoretical slope $1$, logarithms are decimal.}
\label{fig4}
\end{figure}
The above algorithm is optimal for not very large times $t$.  If one
is interested in simulating the dynamics on the attractor for
large $t$, then the size of the garbage register can be significantly reduced.
Indeed, at any $t$, copying the results $x(t),y(t)$ in two 
additional registers
and reversing the sequence of gates allows to erase the garbage
and reproduce $x_0,y_0$.  This 
procedure can be done recursively following the strategy of ``reversible
pebble game'', the description of which can be found in \cite{preskill}.
In the simplest version, $n_t$ qubits in the garbage register
(plus the two additional registers for $x_0,y_0$)
allow to perform map iterations up to $t \sim 2^{n_t}$.  This
gives only a polynomial increase in the number of elementary
quantum operations, being proportional to $n_g^{1.58}$.  The procedure
becomes cost-effective for $t \gg n_q$.

The algorithm described above can be generalized to other dissipative 
maps.  For example, modular multiplications can be performed in
$O(n_q^2)$ operations,
as described in \cite{barenco}. This allows to simulate efficiently
the map (\ref{map}) with $x^2$ term in first equation and also
the H\'enon attractor
$\bar{x}=y+1-ax^2, \;\bar{y}=bx$ \cite{lichtenberg}.  Such an algorithm
can also be adapted to perform the finite-step integration of the Lorenz
system $\dot{x}=-\sigma (x -y), \; \dot{y}= -xz + rx -y, \; 
\dot{z}= xy -bz$ \cite{lorenz}.  However the simulation of the dissipative
dynamics of these models requires more qubits than for (\ref{map}). 

Thus on the example of the map (\ref{map}), we have shown that a 
quantum computer can efficiently simulate dissipative irreversible dynamics.
A quantum processor with 70 qubits will be able to
provide new information about small-scale structures of strange attractors
inaccessible to modern supercomputers.

We thank CalMiP in Toulouse and IDRIS in Orsay 
for access to their supercomputers which were used to simulate 
quantum computations.
This work was supported in part by the EC  RTN contract HPRN-CT-2000-0156
and also by the NSA and ARDA under ARO contract No. DAAD19-01-1-0553.

\end{document}